\documentclass[
    ,final            
  ]
  {aipproc}

\layoutstyle{6x9}

\def \sect #1 {\setcounter{equation} 0\section{#1}}
\def \be  {\begin{equation}}
\def \ee  {\end{equation}}
\def \ba  {\begin{eqnarray}}
\def \ea  {\end{eqnarray}}
\def \baa {\begin{eqnarray*}}
\def \eaa {\end{eqnarray*}}
\def \bb  {\begin{thebibliography}}
\def \eb  {\end{thebibliography}}

\def \lab #1 {\label{#1}}

\def \fracs #1#2 {\mbox{\small $\frac{#1}{#2}$}}

\def \bin #1#2 {{\left({#1}\atop{#2}\right)}}
\def \as {\relax\ifmmode\alpha_s\else{$\alpha_s${ }}\fi}

\def \al #1 {\frac {\as({#1})}{\pi} }
\def \ds #1 {\ooalign{$\hfil/\hfil$\crcr$#1$}}

\newcommand \bea{\begin{eqnarray}}
\newcommand \eea{\end{eqnarray}}

\def\hepph  #1 {{\tt hep-ph/#1}}

\begin{document}

\begin{flushright}
YITP-SB-05-50
\end{flushright}

\title{Fragmentation, NRQCD and Factorization 
in  Heavy Quarkonium Production}


\author{Gouranga C.\ Nayak}{
  address={C.N.\ Yang Institute for Theoretical Physics,
Stony Brook University, SUNY, 
Stony Brook, New York 11794-3840, U.S.A.}
}

\author{Jian-Wei Qiu}{
  address={Department of Physics and Astronomy,
Iowa State University, 
Ames, Iowa 50011-3160, U.S.A.}
}

\author{George Sterman}{
  address={C.N.\ Yang Institute for Theoretical Physics,
Stony Brook University, SUNY,
Stony Brook, New York 11794-3840, U.S.A.}
}

\begin{abstract}
We discuss factorization in heavy quarkonium production in
high energy collisions using NRQCD. Infrared divergences at NNLO
are not matched by conventional NRQCD matrix elements.
However, we show that gauge invariance
and factorization require that conventional NRQCD
production matrix elements be modified to include
Wilson lines or non-abelian gauge links.
With this modification NRQCD factorization
for heavy quarkonium production is restored at NNLO.
\end{abstract}

\maketitle


\section{Introduction}

Quarkonium production and decay have
been the subject of a vast theoretical literature and of intensive 
experimental study \cite{review},
in which the effective field theory nonrelativistic QCD
(NRQCD) \cite{bodwin94} has played a guiding role.
NRQCD offers a systematic formalism to separate
dynamics at the perturbative mass
scale of the heavy quarks
from nonperturbative dynamics, through an expansion
in relative velocity within the pair forming the
bound state. 
An  early success of
NRQCD was to provide a  framework for
the Tevatron Run I data
on high-$p_T$  heavy quarkonium production \cite{teva},
and it has been extensively applied to heavy quarkonia
in both collider and fixed target experiments \cite{review}.

In contrast to quarkonium decay, fully convincing arguments
have not yet been given for NRQCD factorization
as it is applied to high-$p_T$ production processes.
Here we review new tests of NRQCD factorization for high-$p_T$ quarkonia production
at next-to-next-to-leading order (NNLO) \cite{nqs}.
At NNLO we find infrared divergences that do not fall precisely into
the pattern suggested in Ref.\ \cite{bodwin94}.
These divergences may, however, be incorporated into
color octet matrix elements by a technical redefinition in which
Wilson lines which makes them gauge invariant, restore
factorization at NNLO.

If we assume NRQCD factorization, we have 
\begin{equation}
d\sigma_{A+B\to H+X}(p_T) = \sum_n d\hat\sigma_{A+B\to 
c\bar{c}[n]+X}(p_T)\,
\langle {\mathcal  O}^H_n\rangle\, ,
\label{nrfact}
\end{equation}
where the ${\mathcal O}^{H}_n$ are NRQCD operators
for state $H$, which were introduced in Ref.\ \cite{bodwin94} in the form
\ba
{\mathcal O}^H_n(0)
=
\chi^\dagger{\mathcal K}_n\psi(0)\, \left(a^\dagger_Ha_H\right)\,
\psi^\dagger{\mathcal K}'_n\chi(0)\, ,
\label{Ondef1}
\ea
where $a^\dagger_H$ is the creation operator for state $H$, $\chi$ ($\psi$) are
two component Dirac spinors and where
${\mathcal K}_n$ and ${\mathcal K}'_n$ involve products of color
and spin matrices, and at higher dimensions of covariant derivatives.

Our gauge invariant redefinition of production operators
in octet representation is the replacement
(which we refer to as a gauge completion),
\ba
{\mathcal O}^H_n(0)
\to
\chi^\dagger{\mathcal K}_{n,c}\psi(0)\, \Phi_l^\dagger 
[0,A]_{cb}\left(a^\dagger_Ha_H\right)\,
\Phi_l [0,A]_{ba}\, \chi^\dagger {\mathcal K}'_{n,a}\psi(0)\, ,
\label{replace}
\ea
where we have exhibited the color indices of the octets. Here
\ba
\Phi_l [0,A]~=~{\cal P} {\rm exp}[-ig \int_0^\infty d\lambda l \cdot A(\lambda l)]
\ea
is a Wilson line or gauge link, with $l^2$=0. The field $A=A^a\, T^a$ with $T^a$ 
being the generator in the adjoint representation.

Representative diagrams for the fragmentation function at NNLO
are shown in Fig.\ 1 in which the off shell lines are indicated 
by the heavy dots.   The full infrared
fragmentation function is generated by taking all
allowed cuts of the remaining lines of each such diagram.
We are concerned only with diagrams that connect octet to
singlet quark states, and which are not topologically
factorized, since these are the potential sources of
nonfactoring behavior in both the fragmentation function and
related cross sections. The original argument for NRQCD factorization
was based on the conjecture that all infrared
regions in these diagrams cancel after this limited
sum over cuts \cite{bodwin94}.
In fact, this is
the case at NNLO only if we employ the
gauge-completed definitions for NRQCD
matrix elements, as in Eq.\ (\ref{replace}) above.

\begin{figure}
  \includegraphics[width=.3\columnwidth,height=.2\textheight]{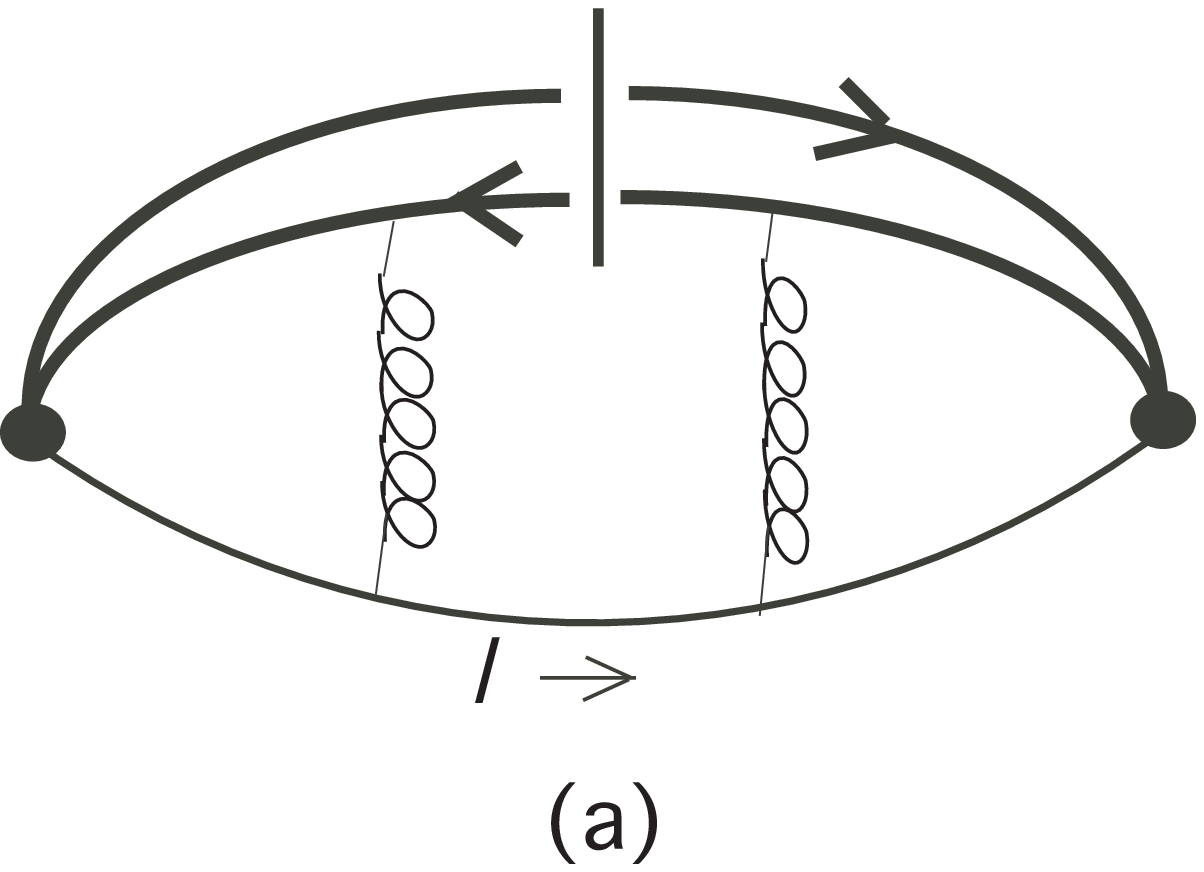}
  \includegraphics[width=.3\columnwidth,height=.2\textheight]{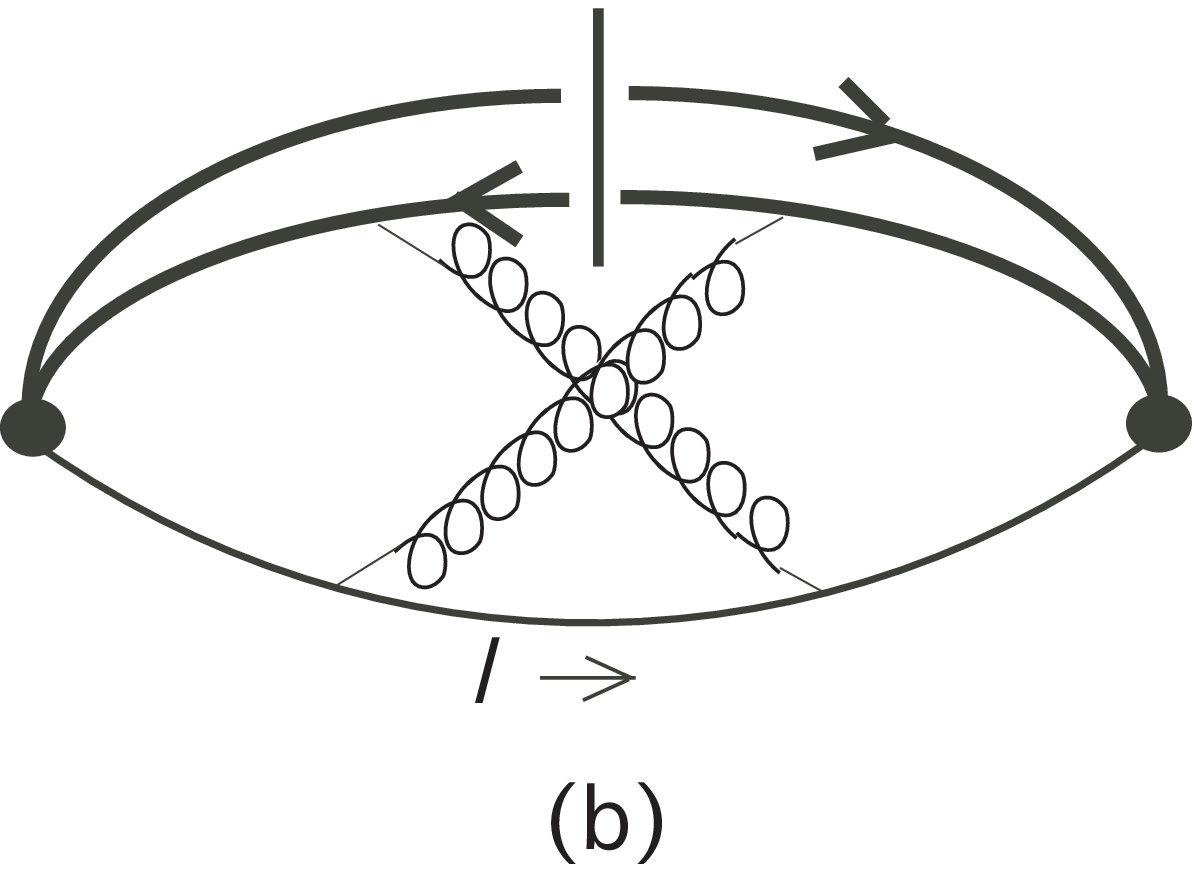}
  \includegraphics[width=.3\columnwidth,height=.2\textheight]{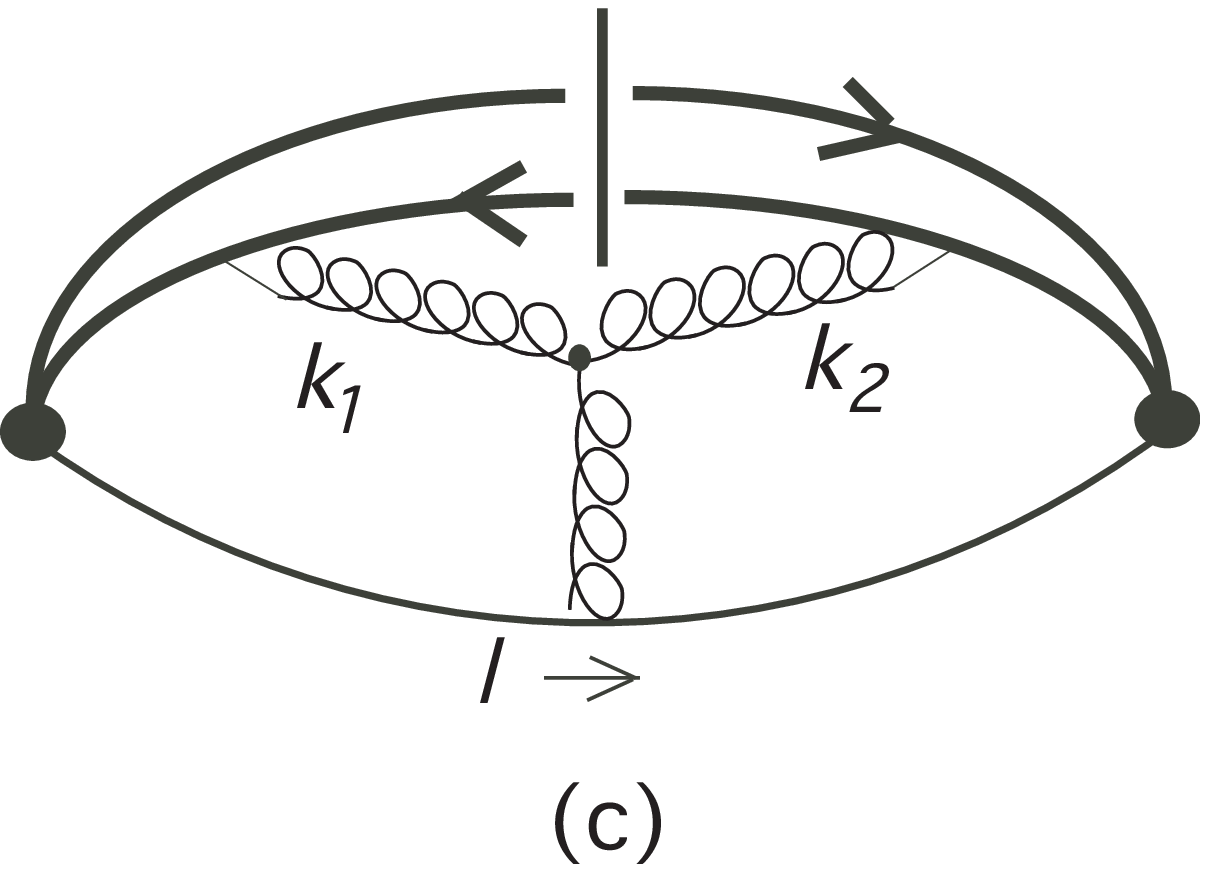}
\label{fig2}
\caption{Representative NNLO contributions to $g\to c\bar{c}$ 
fragmentation in eikonal approximation.}
\end{figure}

The individual classes of
diagrams in Fig.\ 1a and 1b, for which
two gluons are exchanged between the
quarks and the Wilson line, satisfy the infrared
cancellation conjecture of Ref.\ \cite{bodwin94}, by summing over the
possible cuts and connections to quark and
antiquark lines, as do diagrams that have three gluon-eikonal
vertices on the quark pair and one on the  Wilson line.  For the class 
of diagrams
related to Fig.\ 1c, however, with a three-gluon interaction, this 
cancellation
fails. Expanding to second order
in the relative momentum $q$, the full
contribution from Fig.\ 1c, found by cutting the gluon line
$k_1$ and the Wilson line,  can be written as

\newpage

\begin{eqnarray}
&&\Sigma^{(2)}(q,l)~
=~ -16i\, g^4\mu^{4\varepsilon}\, \int \frac{d^D k_1}{(2\pi)^D} 
\frac{d^D
k_2}{(2\pi)^D}~2\pi~\delta(k_1^2) \;
l^\lambda\, V_{\nu\mu\lambda} [k_1,k_2] \nonumber\\
&& \hspace{40mm} \times
\ [q^\mu (P\cdot k_1) - (q\cdot k_1) P^\mu] \,
  ~ [q^\nu (P\cdot k_1)-(q\cdot k_2) P^\nu]~  \nonumber \\
&& \hspace{40mm} \times \frac{1}{[P\cdot k_1 +i\epsilon]^2~
[P\cdot k_2 - i\epsilon ]^2} \nonumber\\
&& \hspace{40mm} \times \frac{1}{[k_2^2 - i \epsilon]~[(k_2- k_1)^2 - 
i\epsilon]~
  [l\cdot (k_1 - k_2) - i\epsilon]}
\, ,
\label{nnlopt}
\end{eqnarray}
where $V_{\nu\mu\lambda} [k_1,k_2]$ represents the
momentum part of the three-gluon vertex. Summing over all contributions, 
however, we find a noncancelling real infrared pole in the fragmentation 
function (in the quarkonium rest frame $\vec{P}$ =0) \cite{nqs}, 
\begin{equation}
\Sigma(P,q,l)~=~\alpha_s^2~\frac{4}{3 \varepsilon}~\frac{\vec{q}\, 
{}^2}{4m_c^2}
~=~\alpha_s^2~\frac{1}{3 \varepsilon}~\frac{\vec{v}\, {}^2}{4}\, ,
\label{gn1}
\end{equation}
where $\vec{v}$ is the relative velocity of the heavy quark pair.

Eq.\ (\ref{gn1}) shows explicitly the breakdown of topological
factorization of infrared dependence at NNLO.
Its presence implies that
  infrared poles would appear in coefficient functions at NNLO
and beyond when the factorization is carried out with octet
NRQCD matrix elements defined in the conventional manner,
Eq.\ (\ref{Ondef1}).  On the other hand, when defined according
to its gauge-completed form (\ref{replace}), each octet
NRQCD matrix element itself generates precisely the
same pole terms given in (\ref{gn1})  above.
Thus, at least at NNLO and  to order $v^2$,
when the non-perturbative NRQCD matrix are defined 
according to Eq. (\ref{replace}), the factorization is restored.
We are investigating the extensions of this result.


\begin{theacknowledgments}
This work was supported in part
by the National Science Foundation, grants PHY-0071027, PHY-0098527 and 
PHY-0354776,
and by the Department of Energy, grant DE-FG02-87ER40371.
\end{theacknowledgments}



\bibliographystyle{aipproc}   


\end{document}